# Variational Quantum Search with Shallow Depth for Unstructured Database Search

Junpeng Zhan[1*]

ABSTRACT

With the advent of powerful quantum computers, the quest for more efficient quantum algorithms becomes crucial in attaining quantum supremacy over classical counterparts in the noisy intermediate-scale quantum era. While Grover's search algorithm and its generalization, quantum amplitude amplification, offer quadratic speedup in solving various important scientific problems, their exponential time complexity limits scalability as the quantum circuit depths grow exponentially with the number of qubits. To overcome this challenge, we propose Variational Quantum Search (VQS), a novel algorithm based on variational quantum algorithms and parameterized quantum circuits. We show that a depth-10 Ansatz can amplify the total probability of $k$ ($k \geq 1$) good elements, out of $2^n$ elements represented by $n+1$ qubits, from $k/2^n$ to nearly 1, as verified for $n$ up to 26, and that the maximum depth of quantum circuits in the VQS increases linearly with the number of qubits. Our experimental results have validated the efficacy of VQS and its exponential advantage over Grover's algorithm in circuit depth for up to 26 qubits. We demonstrate that a depth-56 circuit in VQS can replace a depth-270,989 circuit in Grover's algorithm. Envisioning its potential, VQS holds promise to accelerate solutions to critical problems.

## 1. INTRODUCTION

A vast amount of data is generated every day, and the pace of data generation is increasing fast. Most data, such as text, image, voice, video, is unstructured. Consequently, the ability to search through unstructured databases [1] has become crucial in numerous domains including big data analytics, social media analysis, natural language processing, healthcare and life sciences, the internet of things, e-commerce, and recommendation systems. Traditional algorithms typically require searching through half the elements of a database, on average, to locate a specific target element (called as *good element* below). However, the emergence of quantum computing has opened up new possibilities.

A review of different quantum search algorithms is available in Ref. [2]. Among these algorithms, Grover's search algorithm (GSA) [3], [4] is widely known for its quadratic speedup in searching unstructured databases. Pure quantum algorithms can be classified into two types based on whether they rely on GSA or Shor's algorithm [5], underscoring the scientific and theoretical significance of GSA. Quantum amplitude amplification (QAA) [6], [7] is a generalization of GSA. Both GSA and QAA offer quadratic speedup for various important problems such as searching unstructured databases [5], [8], [9], NP-complete problems [5], [10]–[13], quantum state preparation [14]–[16], quantum counting [17]–[19], factoring [5], [20], [21], collision problem [22], machine learning [23]–[27], and more. Many variants [13], [28]–[32], implementation [33]–[38], and applications [2], [5], [10] of GSA have been extensively investigated. Ref. [29] proposed a variational learning Grover's quantum search algorithm that exhibits improvement over GSA for 3- and 4-qubits scenarios, but lacks advantages for larger qubit cases. GSA has been considered as the fastest possible quantum search of an unstructured database [5], [9], [39]–[41]. However, QAA, GSA, and GSA's variants mentioned above, suffer from exponentially increasing circuit depth as the number of qubits grows [39], [41].

To address this challenge, we propose *Variational Quantum Search (VQS)*, a novel Variational Quantum Algorithm (VQA) [42]–[45] that leverages Noisy Intermediate-Scale Quantum (NISQ) processors. VQAs, including VQS, show promising potential in achieving quantum supremacy over classical computing by utilizing fewer qubits and lower circuit depth compared to pure quantum algorithms [42], [43], [46]. We employ an iterative process between classical and quantum computers to update the Ansatz of the VQS, as depicted in Fig. 1. We have devised an objective function for the optimizer in VQS, guaranteeing that the *global minimum* objective function is linked to both the optimal parameters for the Ansatz and the total probability of good elements being amplified to 1. Through our experiments, we demonstrate VQS's capability to significantly amplify the total probability of one or multiple good elements.

[1] Department of Renewable Energy Engineering, Alfred University, Alfred, NY 14802, USA.
*corresponding author: JZ.   Email: zhanj@alfred.edu

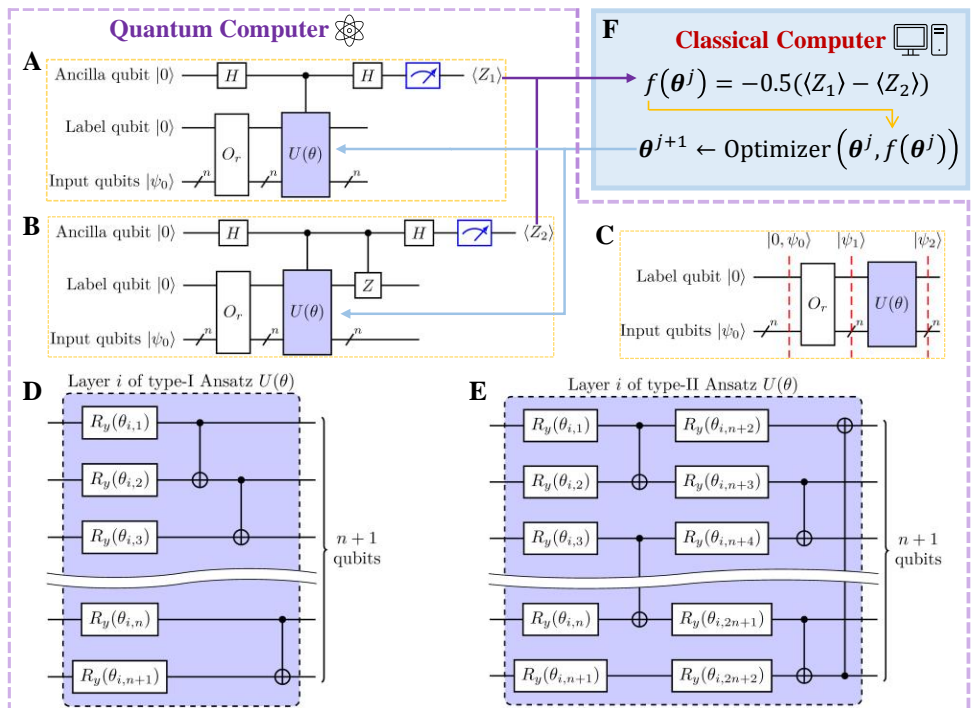

**Fig. 1. Schematic of VQS.** VQS uses an iterative process between (**A**),(**B**) and (**F**) to find the optimal parameters of Ansatz. (**A**),(**B**), two quantum circuits used in VQS, respectively. (**C**), a parameterized quantum circuit that is executed once following the final iteration of the VQS, using the parameters determined by the last iteration. (**D**),(**E**), a layer of two types of Ansatzs, $U(\boldsymbol{\theta})$, respectively. (**F**), the classical part of VQS. The notation with a forward slash and '*n*' in the upper right corner indicates *n* qubits. In the $j^{\text{th}}$ iteration of VQS, the measurement expectations $\langle Z_1 \rangle$ and $\langle Z_2 \rangle$ from (**A**) and (**B**), respectively, are sent to a classical computer (**F**), which calculates the objective $f(\boldsymbol{\theta}^j)$ and new Ansatz parameters $\boldsymbol{\theta}^{j+1}$. Then $\boldsymbol{\theta}^{j+1}$ is then used in the $(j+1)^{\text{th}}$ iteration.

This paper presents two types of low-depth *Ansatzs* and verifies that a depth-10 Ansatz can effectively amplify the total probability of $k$ ($k\geq 1$) good elements out of $2^n$ elements represented by $n+1$ qubits, from $k/2^n$ to nearly 1. This has been confirmed for values of *n* up to 26. This paper also details that the maximum depth of quantum circuits in VQS grows linearly with the number of qubits. Our experimental results confirm the VQS' *exponential advantage* over GSA in circuit depth for up to 26 qubits. As an example, this paper shows that a depth-56 circuit in VQS can replace a depth-270,989 circuit in GSA, showcasing that VQS is particularly suitable for NISQ computers.

2. METHODS
2.1 Problem description
Consider a Boolean function $\chi : X \rightarrow \{0,1\}$ that partitions set *X* between its *good* and *bad* elements, where *x* is considered good if $\chi(x)=1$ and bad otherwise[6]. Consider a normalized state vector $|\psi_0\rangle = \sum_{x \in X} \alpha_x |x\rangle$ which is a quantum superposition of the elements of *X*. Let *a* denote the probability that a good element is obtained if $|\psi_0\rangle$ is measured. QAA is a process that can find a good *x* using $O(1/\sqrt{a})$ applications of a unitary operator [6]. GSA is a special instance of QAA where $|\psi_0\rangle$ is an equal superposition of all members of *X* (i.e., $\alpha_x$ is the same for all *x*). GSA uses an oracle to add a negative phase to the good element.

## 2.2 Variational Quantum Search (VQS)

Our VQS has three quantum circuits (Fig. 1A-C) in its quantum part, while using an optimizer in the classical part. The interaction between these two parts is like a general VQA. In the classical part (Fig. 1F), an optimizer is used to determine new values of parameter $\boldsymbol{\theta}^{j+1}$, which are then fed into the Ansatz $U(\boldsymbol{\theta})$ depicted in Fig. 1A,B for the next iteration. In each iteration, both Ansatzs in Fig. 1A,B use the same parameters obtained from the optimizer's output (see Fig. 1F). Note that Fig. 1A,B is involved in the iterative process of VQS, while Fig. 1C runs only once after the final iteration. In this paper, we employ ADAM [47] as the optimizer, although alternative optimizers could also be used [48], [49].

Fig. 1A-C have identical Ansatz structures, wherein an Ansatz comprises one or more layers connected in series. Different layers are the same except that their parameters can be different. We investigate two types of Ansatzs in this paper. Each layer of type-I Ansatz (Fig. 1D) has a depth of $n+1$, consisting of one layer of $R_y$ gates and CNOT gates sequentially applied on every two adjacent qubits. Each layer of type-II Ansatz (Fig. 1E) has a depth of 5, consisting of CNOT gates alternatively applied on two adjacent qubits with two $R_y$ gates before each CNOT gate. In this paper, we maintain the separation of type-I (Fig. 1D) and type-II (Fig. 1E) structures, although it is possible to combine them into a single Ansatz.

In all quantum circuits shown in this paper, we use the convention that the top wire affects the most significant (leftmost) qubit. In Fig. 1A,B, we use Pauli-Z measurement at the top wire.

As shown in Fig. 1A-C, each quantum circuit has a label qubit and uses an **oracle**, $O_r$, to set it to $|1\rangle$ and $|0\rangle$ for good and bad elements, respectively:

$$\begin{cases} O_r|0\rangle|x\rangle = |1\rangle|x\rangle & \text{for good elements, that is, } \chi(x) = 1 \\ O_r|0\rangle|x\rangle = |0\rangle|x\rangle & \text{for bad elements, that is, } \chi(x) = 0 \end{cases} \quad (1)$$

In this paper, we use multi-controlled single target CNOT gate as the oracle.

The **objective function**, denoted as $f(\boldsymbol{\theta})$, is defined as $f(\boldsymbol{\theta}) = -0.5(\langle Z_1 \rangle - \langle Z_2 \rangle)$. We show below that this function ensures that when $f(\boldsymbol{\theta})$ is minimized, the state $|\psi_2\rangle$ is exclusively a superposition of good elements, and the probability of measuring bad elements is 0.

In Fig. 1C, the Ansatz acts on state $|\psi_1\rangle$ to obtain output state $|\psi_2\rangle$, which can be expressed as:

$$|\psi_2\rangle = U(\boldsymbol{\theta})|\psi_1\rangle = U(\boldsymbol{\theta})O_r|0, \psi_0\rangle \quad (2)$$

For the convenience of analysis, we use subscripts b and g to indicate bad and good elements, respectively, and write the state $|\psi_0\rangle$ in a vector form:

$$|\psi_0\rangle = \left[\alpha_{b_1}, \alpha_{b_2}, \cdots, \alpha_{b_{N_b}}, \alpha_{g_1}, \alpha_{g_2}, \cdots, \alpha_{g_{N_g}}\right]^T \quad (3)$$

where $N_b$ and $N_g$ are the number of bad and good elements, respectively, with $N_b + N_g = N = 2^n$, where $n$ denotes the number of input qubits. For the sake of simplicity in the expression, we assume that the good elements are positioned at the end of the vector $|\psi_0\rangle$. However, their location can be anywhere within the vector and does not need to be contiguous. Then we have

$$|0, \psi_0\rangle = [\underbrace{\alpha_{b_1}, \alpha_{b_2}, \cdots, \alpha_{b_{N_b}}, \alpha_{g_1}, \alpha_{g_2}, \cdots, \alpha_{g_{N_g}}}_{\text{1st half: } N \text{ elements}}, \underbrace{0, 0, \cdots, 0}_{\text{2nd half: } N \text{ elements}}]^T \quad (4)$$

$$|\psi_1\rangle = O_r|0, \psi_0\rangle = [\underbrace{\alpha_{b_1}, \alpha_{b_2}, \cdots, \alpha_{b_{N_b}}, 0, \cdots, 0}_{\text{1st half: } N \text{ elements}}, \underbrace{0, \cdots, 0, \alpha_{g_1}, \alpha_{g_2}, \cdots, \alpha_{g_{N_g}}}_{\text{2nd half: } N \text{ elements}}]^T \quad (5)$$

Each of the previous two equations has $N$ zero elements.

$$|\psi_2\rangle = U(\boldsymbol{\theta})|\psi_1\rangle \quad (6)$$

Note that the oracle $O_r$ assigns label $|1\rangle$ to the good elements, resulting in their amplitudes transitioning from the first half of the state vector (as shown in equation 4) to the second half (as depicted in equation 5).

Next, we compute the measurement expectations $\langle Z_1 \rangle$ and $\langle Z_2 \rangle$ from Fig. 1A,B, respectively. Note that the states after the oracle $O_r$ (disregard the ancilla qubit) in all Fig. 1A-C are identical and represented as $|\psi_1\rangle$. Both Fig. 1A,B use the Hadamard Test [50] procedure. For ease of analysis, we express $|\psi_2\rangle$ as:

$$|\psi_2\rangle = [\beta_1, \beta_2, \cdots, \beta_N, \beta_{N+1}, \cdots, \beta_{2N-1}, \beta_{2N}]^T \quad (7)$$

where

$$\sum_{i=1}^{2N}|\beta_i|^2 = 1 \quad (8)$$

according to the normalization condition. Then we have:

$$\langle Z_1 \rangle = \langle \psi_1|U(\boldsymbol{\theta})|\psi_1 \rangle = \langle \psi_1|\psi_2 \rangle \quad (9)$$

$$\langle Z_2 \rangle = \langle \psi_1|Z \otimes I^{\otimes n}U(\boldsymbol{\theta})|\psi_1 \rangle = \langle \psi_1|Z \otimes I^{\otimes n}\psi_2 \rangle \quad (10)$$

$$f(\boldsymbol{\theta}) = -0.5(\langle Z_1 \rangle - \langle Z_2 \rangle) = -0.5\langle \psi_1|(I-Z) \otimes I^{\otimes n}\psi_2 \rangle$$
$$= -\langle \underbrace{[0,\cdots,0,\alpha_{g_1},\alpha_{g_2},\cdots,\alpha_{g_{N_g}}]^T}_{N \text{ elements}}, [\beta_{N+1},\cdots,\beta_{2N-1},\beta_{2N}]^T \rangle \quad (11)$$

where $Z$ and $I$ represent the Pauli-Z gate and the identity matrix, respectively. Note that $0.5(I-Z) \otimes I^{\otimes n} = \text{diag}([0,0,\cdots,0,1,1,\cdots,1])$ where the number of 0 and 1 is $N$ for each, and the notation diag(•) indicates the creation of a diagonal matrix. Notably, equation 11 represents the inner product between the 2$^{\text{nd}}$ half of $|\psi_1\rangle$ and $|\psi_2\rangle$.

2.3 Optimization problem

In the analysis given below, we assume all amplitudes of $|\psi_1\rangle$ and $|\psi_2\rangle$ are real numbers, i.e., $\alpha_i$ and $\beta_i$ are real numbers. We can then solve the following optimization problem to determine the minimum value of $f(\boldsymbol{\theta})$:

$$\text{minimize } f(\boldsymbol{\theta}) = -\sum_{i=1}^{N_g} \alpha_{g_i}\beta_{2N-N_g+i} \quad (12)$$

$$\text{subject to: (8)}$$

where $\alpha_{g_i}$ is a known value from the input state $|\psi_0\rangle$, and $\beta_{2N-N_g+i}$ is the unknown variable. As a result, the objective $f(\boldsymbol{\theta})$ is a bounded linear function. We can convert it into minimizing the Lagrange function $\mathcal{L}(\beta) = -\sum_{i=1}^{N_g} \alpha_{g_i}\beta_{2N-N_g+i} + \lambda(\sum_{i=1}^{2N} \beta_i^2 - 1)$.

An optimal solution $\beta$ must satisfy the Karush-Kuhn-Tucker (KKT) conditions[51] as presented in equations 13-15:

$$\frac{\partial \mathcal{L}}{\partial \beta_i} = 2\lambda\beta_i = 0, \quad \forall i \in \{1,2,\cdots,2N-N_g\} \quad (13)$$

$$\frac{\partial \mathcal{L}}{\partial \beta_{2N-N_g+i}} = -\alpha_{g_i} + 2\lambda\beta_{2N-N_g+i} = 0, \quad \forall i \in \{1,2,\cdots,N_g\} \quad (14)$$

$$\frac{\partial \mathcal{L}}{\partial \lambda} = \sum_{i=1}^{2N} \beta_i^2 - 1 = 0 \quad (15)$$

In equation 14, if $\lambda = 0$, we have $\alpha_{g_i} = 0, \ \forall i \in \{1,2,\cdots,N_g\}$, indicating that the probability of all good elements is 0. However, since we assume that there is at least one good element with a non-zero probability, $\lambda$ must be non-zero. By solving equations 13-14, we can obtain:

$$\begin{cases} \beta_i = 0, & \forall i \in \{1,2,\cdots,2N-N_g\} \\ \beta_{2N-N_g+i} = \alpha_{g_i}/2\lambda, & \forall i \in \{1,2,\cdots,N_g\} \end{cases} \quad (16)$$

Substituting equation 16 into equation 15 gives:

$$\sum_{i=1}^{N_g} \alpha_{g_i}^2/4\lambda^2 = 1 \implies \lambda = \pm 0.5\sqrt{\sum_{i=1}^{N_g} \alpha_{g_i}^2} \quad (17)$$

Substituting equation 17 back into equation 16, we have:

$$\beta_{2N-N_g+i} = \alpha_{g_i}/2\lambda = \pm \alpha_{g_i}/\sqrt{\sum_{i=1}^{N_g} \alpha_{g_i}^2}, \quad \forall i \in \{1,2,\cdots,N_g\} \quad (18)$$

In summary, by solving equations 13-15, we obtain the optimal solution $\beta$, as shown in equation 18 and the first row in equation 16. Since $f(\boldsymbol{\theta})$ is a linear function, it reaches its global minimum, $-\sum_{i=1}^{N_g} \alpha_{g_i} \beta_{2N-N_g+i} = -\sum_{i=1}^{N_g} \alpha_{g_i}^2 / \sqrt{\sum_{i=1}^{N_g} \alpha_{g_i}^2}$, when $\beta_{2N-N_g+i}$ takes the positive sign in equation 18. In other words, $f(\boldsymbol{\theta})$ reaches its global minimum, when $\beta_{2N-N_g+i}$ is proportional to $\alpha_{g_i}$ and satisfies the normalization condition given in equation 8.

By substituting equation 16 into equation 7 and comparing it with equation 5, we see that when $f(\boldsymbol{\theta})$ is minimized, all elements in $|\psi_2\rangle$ become 0 except for the ones at the same positions as $\alpha_{g_i}$ in $|\psi_1\rangle$, $\forall i \in \{1, 2, \cdots, N_g\}$.

This means that during the transition from $|\psi_1\rangle$ to $|\psi_2\rangle$, the amplitudes of all bad elements reduce to 0, the original elements with 0 amplitude remain unchanged, and the amplitudes of the good elements are amplified while maintaining their original ratio and satisfying the normalization condition. This constitutes a perfect quantum search, where $|\psi_1\rangle$ is a superposition of all bad and good elements, and the output state $|\psi_2\rangle$ is a superposition of all good elements with their total probability almost equal to 1. If noise and errors are not considered, measuring $|\psi_2\rangle$ will yield only the good elements. This also means that when the objective $f(\boldsymbol{\theta})$ is minimized, the corresponding $\boldsymbol{\theta}$ achieves its optimal value, resulting in $|\psi_2\rangle$ being exactly the superposition of all good elements with amplified probabilities.

VQS utilizes two termination criteria in its iterative process. The first criterion is reaching a given number of iterations (set to 300 in this paper). The second criterion is the occurrence of consecutive small-change events for a given number of times (set to 5 in this paper). A small-change event is defined as the absolute value of the relative change of objective functions in two consecutive iterations being smaller than a given threshold (set to $1 \times 10^{-4}$ in this paper). The iterative process of VQS terminates when either criterion is met, whichever comes first.

### 2.4 Depth of quantum circuits in VQS

In VQS, each layer of the type-I Ansatz (Fig. 1D) for an $n$-qubit input state has a depth of $n$+1. The controlled layer, which introduces control to each gate in the layer, has a depth of $2n$+1. The oracle $O_r$ in Fig. 1A-C can be realized as an $n$-qubit-controlled one-target CNOT gate, denoted as $C^n(X)$. This gate can be decomposed into a circuit of depth $2n-1$, comprising a CNOT gate and ($2n$-2) Toffoli gates (as detailed in Fig. S3). By summing up the depths of each component, we can determine the depths of the circuits depicted in Fig. 1A-C (using a 3-layer type-I Ansatz) to be $8n$+3, $8n$+4, and $5n$+2, respectively.

As an example, consider the case where $n$=26. In Fig. 1C, the oracle is a $C^{26}(X)$ gate, and a one-layer type-II Ansatz with a depth of 5, as shown in Fig. 1E, is employed. The $C^{26}(X)$ gate can be decomposed into a circuit of depth 51, consisting of a CNOT gate and 50 Toffoli gates, according to Fig. S3. Thus, when $n$=26, the circuit illustrated in Fig. 1C has a depth of 56.

### 2.5 Experiment setting

The results presented in this paper were obtained from quantum simulators using Pennylane's devices[52]. Specifically, the results for 2-, 8-, and 14-qubit (20- and 26-qubit) input states were obtained using Pennylane's default.qubit (lightning.gpu) device on an Intel i5-6500 CPU (A40x4 48-GB GPU). The initial values of the parameters $\boldsymbol{\theta}$ in the Ansatz $U(\boldsymbol{\theta})$ were randomly sampled from a uniform distribution between 0 and $2\pi$.

## 3. RESULTS

The goal of VQS, QAA, and GSA is the same: to find the good element(s) in a quantum input state $|\psi_0\rangle$ (see METHODS). To assess the efficacy and stability of the VQS using random initial parameters (as described in METHODS), we independently execute the VQS 100 times for each case. The results are presented in the box plots below. Each box shows the 0th, 25th, 50th, 75th, and 100th percentiles, while outliers are indicated by circles. Further details regarding the experimental settings are given in the METHODS section. To showcase the efficacy and advantages of the VQS, the following sub-sections discuss its scalability and stability, two types of Ansatzs, a comparison with GSA, and its performance in amplifying multiple good elements.

## 3.1 Scalability and stability

To assess its effectiveness and *scalability*, we run the VQS with type-I Ansatz (Fig. 1D), for an *n*-qubit input state where *n*=2, 8, 14, 20, and 26. In this sub-section, we initialize each element with equal probabilities by applying the Hadamard gate to each input qubit, represented as $|\psi_0\rangle = H^{\otimes n}|0\rangle^{\otimes n}$. We use a multi-control CNOT gate to realize the oracle $O_r$. We set the number of good elements to 1 and the number of layers in the Ansatz to 3. The results, presented in Fig. 2A,B, indicate the consistent and excellent performance of the VQS across all runs with random initial parameters, demonstrating its remarkable *stability*.

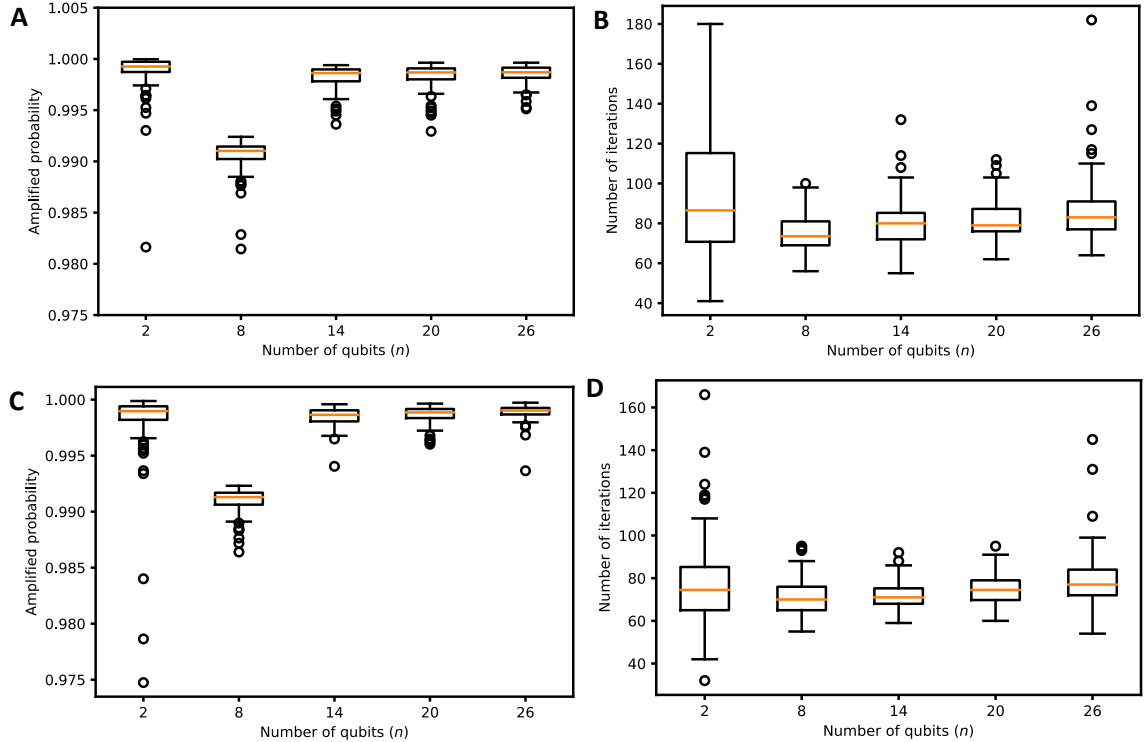

**Fig. 2. Box plot results of the VQS for sole good element.** The VQS uses the 3-layer type-I Ansatz (**A**),(**B**) and the 2-layer type-II Ansatz (**C**),(**D**) to find the sole good element in an *n*-qubit input state, *n*=2,8,14,20, and 26. The VQS runs 100 times independently. (**A**),(**C**), The amplified probability of the sole good element exceeds 0.975 for all runs except for one run in the 3-qubit and the 26-qubit cases in panel (**A**), with probabilities of 0.93 and 0.50, respectively (not displayed for better visibility of the boxes). (**B**),(**D**), The number of iterations required to meet the termination. The median values (represented by red lines) for different cases range between 70 and 90 iterations. The variance is higher in the 2- and 26-qubit cases but smaller in the other three cases.

Moving on to the *scalability* of the VQS, Fig. 2A,B depicts the successful amplification of the probability of the good element. The initial probabilities, calculated as 0.25, $3.9 \times 10^{-3}$, $6.1 \times 10^{-5}$, $9.5 \times 10^{-7}$, and $1.5 \times 10^{-8}$ (corresponding to $1/2^n$), are elevated to nearly 1 for the respective cases. An amplified probability of 1 signifies that a single measurement of the output state ($|\psi_2\rangle$ in Fig. 1C) can identify the sole good element out of a total of $2^n$ elements. For instance, when *n*=26, a single measurement of the 27 qubits can find the sole good element among $2^{26}$, which amounts to $6.7 \times 10^7$ elements. Additionally, Fig. 2B,D demonstrates that the VQS requires a comparable number of iterations across different cases. In summary, while the original probability of the good element decreases *exponentially* with the number of qubits, the VQS consistently amplifies the probability to almost 1, highlighting its impressive *scalability*. In conclusion, the VQS exhibits excellent scalability and stability.

## 3.2 Structure of ansatz and the number of layers

Here, we compare the efficiency of VQS using the type-I (Fig. 1D) and type-II (Fig. 1E) Ansatzs. The results of VQS with 2-layer type-II Ansatz are given in Fig. 2C,D, demonstrating comparable performance to the VQS with a 3-layer type-I Ansatz, as shown in Fig. 2A,B. Conversely, the results of VQS with 1-layer and 2-layer type-I Ansatzs, as well as the 1-layer type-II Ansatz, exhibit significant variance across different runs (see Fig. S1). Therefore, the type-I and type-II Ansatzs should have 3 layers and 2 layers, respectively.

Next, we compare the circuit depth of the controlled $U(\boldsymbol{\theta})$ (used in Fig. 1A,B) rather than $U(\boldsymbol{\theta})$ (used in Fig. 1C), as the former has a deeper structure and determines the maximum circuit depth involved in VQS. The depths of the controlled $U(\boldsymbol{\theta})$ in the 3-layer type-I and 2-layer type-II Ansatzs are $6n+3$ and $6n+6$, respectively. Hence, we recommend employing the 3-layer type-I Ansatz due to its slightly lower depth compared to the 2-layer type-II Ansatz. Furthermore, the circuit depths of both type-I and type-II Ansatzs *increase linearly* with the number of qubits.

### 3.3 Depth of VQS and Grover's algorithm

This sub-section compares the maximum depths of the quantum circuits used by VQS and GSA for solving the same problem discussed in the previous sub-section. The maximum depth of the circuits employed in VQS is $8n+4$ (see METHODS). The circuit depth of GSA is provided in Table S1. The comparison of circuit depth between the two algorithms is depicted in Fig. 3, clearly illustrating that the circuit depth of VQS exhibits *linear growth*, while that of GSA increases *exponentially*. In other words, VQS achieves an exponential advantage over GSA in terms of circuit depth for up to 26 qubits.

To further highlight the advantage of VQS over GSA, here we provide an example of depth comparison. Although the performance of VQS using a 1-layer type-II Ansatz is not as stable as when using a 2-layer Ansatz, it can still amplify the probability of the good element to nearly 1 in most runs for the 8-, 14-, 20-, and 26-qubit cases (see Fig. S1). For instance, in the 26-qubit case, the circuit depicted in Fig. 1C, with a depth of 56 (see METHODS), can amplify the probability of the good element from $1/2^{26} = 1.5 \times 10^{-8}$ to almost 1. On the other hand, GSA requires a depth of 270,989 (see Table S1) to amplify the sane original probability to 0.9. This example underscores the significant advantage of VQS over GSA in terms of circuit depth.

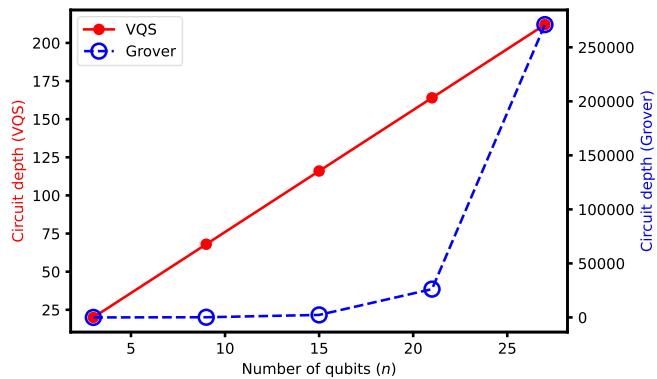

**Fig. 3. Circuit depths: VQS vs GSA.** The circuit depth of the VQS is determined by the circuit depicted in Fig. 1B, which possesses the highest depth among all circuits used in the VQS. The VQS and GSA amplify the probability of the sole good element in an *n*-qubit input state to 1 and 0.9, respectively.

### 3.4 Good elements with different probabilities

Here we demonstrate that the VQS can amplify multiple good elements with varying probabilities, exhibiting a capability similar to that of the QAA. We evaluate the VQS using an *n*-qubit input state, where *n* takes values of 2, 8, 14, 20, and 26. Within this input state, we consider three good elements with a probability ratio of 0.1:0.3:0.6. Initially, the input state $|\psi_0\rangle$ has equal probabilities for all elements. Subsequently, the probabilities of the last three elements, assumed to be the good elements, are adjusted to match the given ratio while maintaining their total probability. Note that the number, location, and probability ratio of good elements can be modified, although specific results are not provided here.

The outcomes, depicted in Fig. 4, demonstrate that the VQS efficiently amplifies the probabilities of the three good elements simultaneously. As a result, their total probability approaches 1 while the probability ratio remains unchanged. These findings indicate that the VQS is highly effective and demonstrates excellent scalability in amplifying multiple good elements, for input states with up to 26 qubits.

A few outliers can be observed at the lower end of Fig. 4D, indicating that in approximately 2 to 3 runs out of 100, the total amplified probability of the good elements is close to 0.5. This implies that, in these instances, the measurement of the output would yield an equal distribution of good and bad elements at 50% each. However, this issue can be addressed by executing the VQS multiple times and measuring the output of each run. By doing so, we can achieve a high probability of obtaining the good elements, such as 98.5% to 99% for the 14-, 20-, and 26-qubit cases illustrated in Fig. 4.

In summary, the VQS demonstrates favorable stability when searching for multiple good elements, although it is slightly less efficient compared to the scenario where only one good element is involved, as discussed in the preceding three sub-sections.

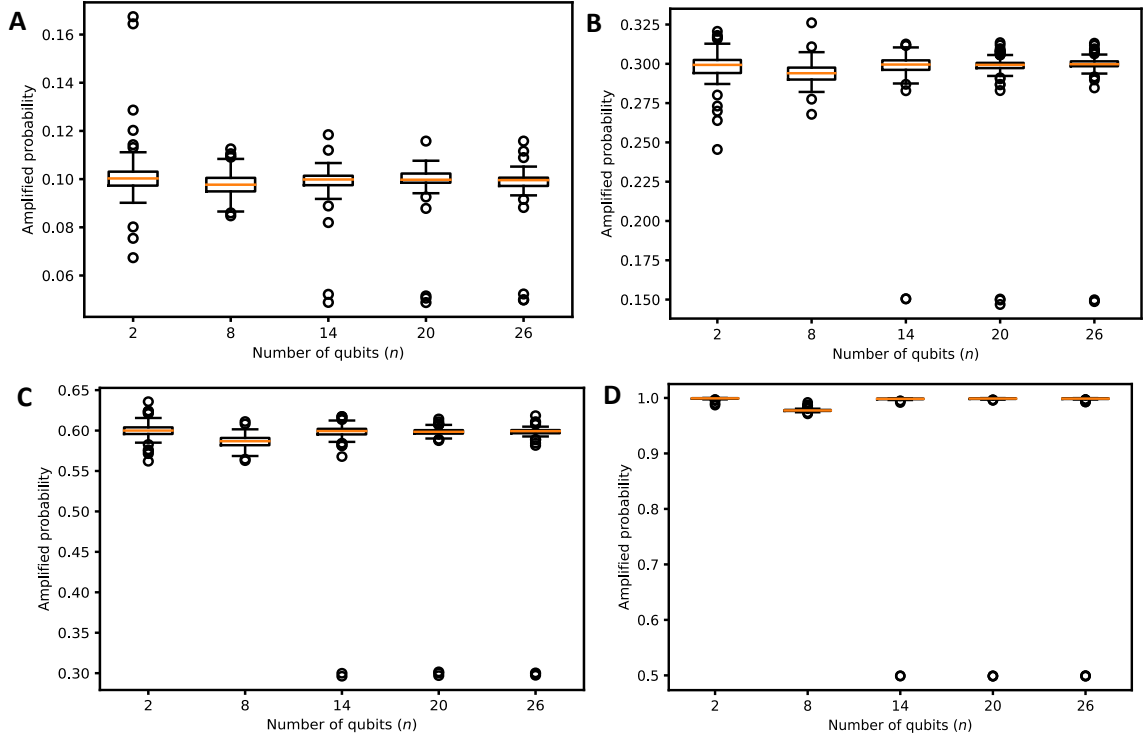

**Fig. 4. Box plot results of the VQS for three good elements.** The VQS uses the 3-layer type-I Ansatz to find the good elements in an $n$-qubit input state, $n$=2, 8, 14, 20, and 26. The VQS runs 100 times independently. (**A**), (**B**), (**C**), amplified probabilities for the good elements, respectively. (**D**), the sum of amplified probabilities of the three good elements. A red line indicates the median value out of 100 probabilities, close to 0.1, 0.3, 0.6, and 1.0 in the four panels, respectively.

5. DISCUSSION AND CONCLUSION

We have thoroughly examined and confirmed the efficiency and superiority of the VQS in finding good element(s) within an input state. The VQS has demonstrated an exponential advantage over GSA, showcasing its scalability up to 26 qubits. Notably, the circuit depth of the VQS increases linearly with the number of qubits, further emphasizing its efficiency. Furthermore, we have showcased the effectiveness of two distinct and efficient shallow-depth Ansatz structures, underscoring the fact that the VQS is not heavily reliant on a specific Ansatz structure. This opens up possibilities for exploring alternative efficient structures [53] in the future. Moreover, we have validated the efficacy of the VQS in searching for either one or multiple good elements. In essence, the VQS successfully accomplishes the same task as both GSA and QAA, further highlighting its versatility and capability.

The VQS has demonstrated an exponential advantage, in circuit depth, over GSA in searching unstructured databases for up to 26 qubits. This finding implies that the VQS may also hold a significant, if not exponential, advantage in solving a wide range of important problems that both GSA and QAA can tackle.

These problems include NP-complete problems [5], [12], quantum state preparation [14], [15], quantum counting [17]–[19], factoring [5], [20], [21], and many others. Based on the extrapolation from the numerical results presented in this paper, we anticipate that the VQS has the potential to excel even beyond 26 qubits. It emerges as an excellent candidate quantum algorithm that surpasses classical algorithms and holds the potential for supremacy in solving complex computational tasks.

In this study, we have successfully demonstrated the effectiveness of the VQS in noise-free simulators. Moving forward, our next objective is to adapt the VQS to noisy environments and validate its performance on real quantum computers. While this paper has established the advantage of the VQS up to 26 qubits, we acknowledge that this limit is due to the memory capacity restriction of the 48-GB GPU used in our experiments. To further advance our research, the next phase will involve numerical verification and mathematical proof of the VQS's scalability for a significantly larger number of qubits. We have already taken an initial step in this direction through the investigation of the VQS's reachability [53] and have compared the performance of different quantum simulators for the VQS [54]. We will employ techniques such as circuit cutting [55], [56] for the numerical verification. These future endeavors will enable us to gain deeper insights into the scalability and supremacy of the VQS.

ACKNOWLEDGMENTS
This research was partially supported by the National Science Foundation (NSF) Engineering Research Initiation (ERI) program, under award number 2138702.

This work used the Delta system at the National Center for Supercomputing Applications through allocation CIS220136 from the Advanced Cyberinfrastructure Coordination Ecosystem: Services & Support (ACCESS) program, which is supported by National Science Foundation grants #2138259, #2138286, #2138307, #2137603, and #2138296.

We acknowledge the use of IBM Quantum services for this work. The views expressed are those of the authors, and do not reflect the official policy or position of IBM or the IBM Quantum team.

APPENDIX
In this appendix, Fig. S1 offers the box plot results of the VQS with reduced layers for both Type-I and Type-II Ansatzes. Additionally, Fig. S2 showcases the Grover iteration, a crucial component within Grover's algorithm. Fig. S3 provides a circuit decomposition for the $C^n(X)$ gate. Algorithm 1 presents the Pseudo code detailing the process of determining the number of Grover iterations required for the execution of Grover's algorithm. Furthermore, Table S1 compiles a tabular representation of the circuit depth utilized by both the VQS and Grover's algorithm across varying numbers of qubits.

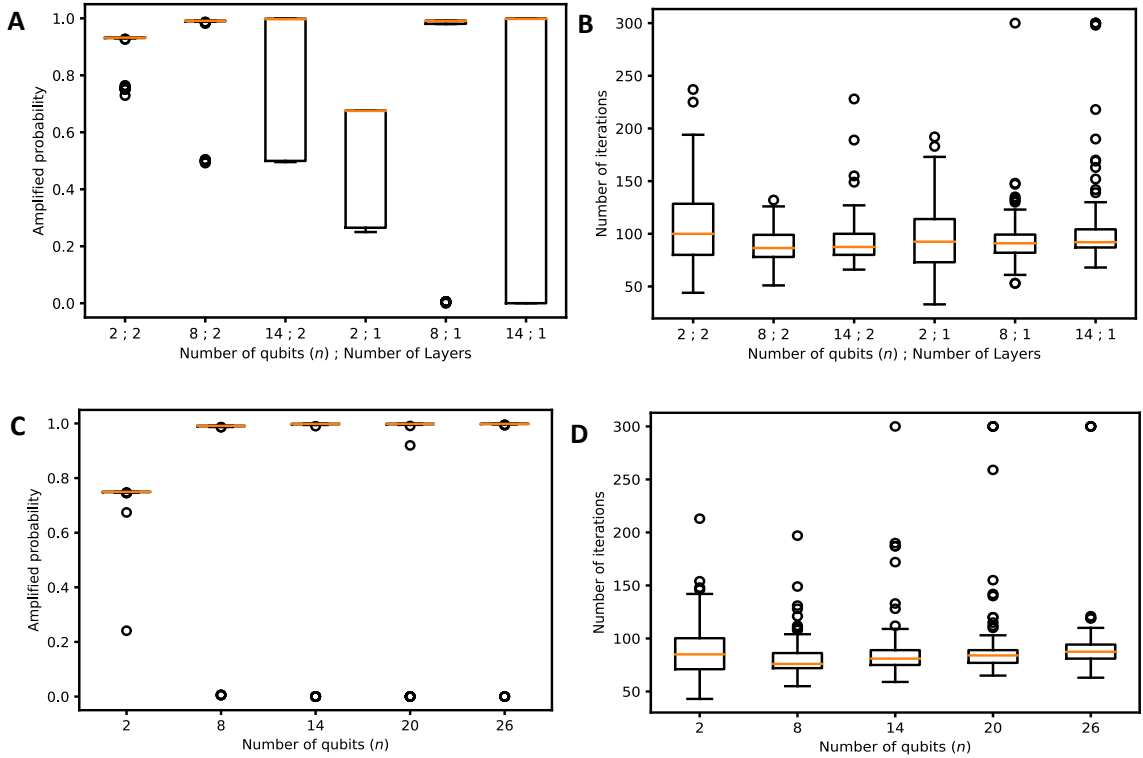

**Fig. S1. Box plot results of the VQS using fewer layers.** The VQS uses 1- and 2-layer type-I Ansatz (**A**),(**B**) and 1-layer type-II Ansatz (**C**),(**D**) to find the sole good element in an *n*-qubit input state, *n*=2,8,14,20, and 26. The VQS runs 100 times independently. (**A**),(**C**), The amplified probability of the good element, where some outliers have low or even 0 values (see the bottom of both panels). (**B**),(**D**), The number of iterations used when a termination criterion is met.

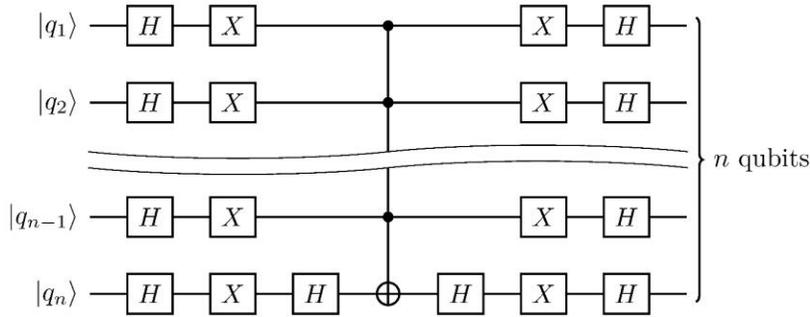

**Fig. S2. Grover iteration.** Grover's algorithm consists of repeated application of the Grover iteration. The (*n*−1)-qubit-controlled one-target CNOT gate, denoted as $C^{n-1}(X)$, has a depth of 2*n*−3 (see Fig. S3). Then the circuit depth of the Grover iteration is 2*n*+1, which is calculated from 2*n*−3+4, where the 4 means the two left-most layers together with the two right-most layers; the two Hadamard gates beside $C^{n-1}(X)$ are not counted (except for the case of *n*=2,3) as each shares a layer with the neighboring Toffoli gate.

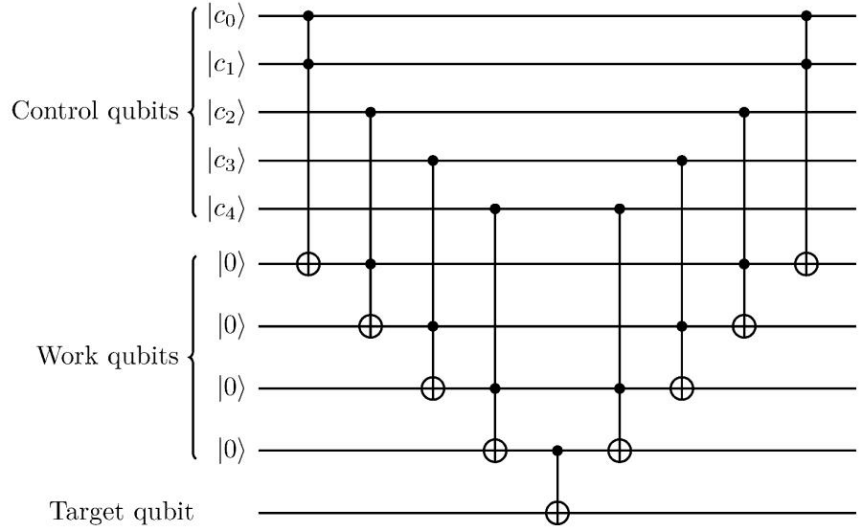

**Fig. S3. Circuit decomposition for $C^n(X)$.** It is decomposed into a depth-$(2n-1)$ circuit consisting of $(2n-2)$ Toffoli gates and a CNOT gate, for the case of $n=5$.

---

**Algorithm 1.** Pseudo code for calculating the number of Grover iterations[†] needed by Grover's algorithm.
---
**Input**: the amplitude of each element, denoted as $\alpha_x$. Let $i=0$.
**Output**: the number of Grover iterations ($n_G$).
1    while $i \leq 1 \times 10^8$
2       Change the sign of the amplitude of good element, i.e., $\alpha_x^g \leftarrow -\alpha_x^g$, where superscript $g$ indicates that element $x$ is a good element.
3       Obtain the average amplitude, denoted as $m$, of all elements, i.e., $m = \sum_{x \in X} \alpha_x / N$.
4       Change the amplitude of element $x$ from $\alpha_x$ to $2m - \alpha_x$.
5       $i \leftarrow i+1$
6       If $(\alpha_x^g)^2 \geq p_s$[‡]: record $n_G = i$ and terminate.

[†]The Grover iteration is given in Fig. S2.
[‡]$p_s$ is the success probability, that is, when the probability of the good element is higher than $p_s$, the while loop in Table S1 terminates.

**Table S1.**

**Circuit depth used by VQS and Grover's search algorithm.**

| $n$ | 2 | 8 | 14 | 20 | 26 |
|---|---|---|---|---|---|
| Circuit depth of VQS ($8n+4$) | 20 | 68 | 116 | 164 | 212 |
| $n_G$ ($p_s=0.5$)* | 1 | 6 | 50 | 402 | 3,215 |
| Circuit depth[†] of Grover's ($p_s=0.5$)[‡] | 7 | 102 | 1,450 | 16,482 | 170,395 |
| $n_G$ ($p_s=0.9$) | 1 | 10 | 80 | 639 | 5,113 |
| Circuit depth of Grover's ($p_s=0.9$) | 7 | 170 | 2,320 | 26,199 | 270,989 |

*The number of Grover iterations ($n_G$) is obtained by running the method described in Table S1.
[‡]The $p_s$ is defined at be bottom of Table S1.
[†]The circuit depth is calculated from ($2n+1$) multiplied by the corresponding $n_G$. For example, the last number 270,989 is obtained from $5,113 \times (2 \times 26+1)$.